\documentclass[
 reprint,
superscriptaddress,
 amsmath,amssymb,
prb,
floatfix,
]{revtex4-2}

\usepackage{graphicx}
\usepackage{dcolumn}
\usepackage{bm}
\usepackage{xcolor}
\usepackage{float}
\usepackage[T1]{fontenc}
\usepackage[colorlinks,urlcolor=blue,citecolor=blue,linkcolor=blue]{hyperref}

\begin{document}

\preprint{APS/123-QED}

\title{Emergent momentum-space topological pseudospin defects in non-Hermitian systems}

\author{Yow-Ming Robin Hu}
 \affiliation{Department of Quantum Science and Technology, Research School of Physics, The Australian National University, Canberra, ACT 2601, Australia}
\author{Elena A. Ostrovskaya}%
\affiliation{Department of Quantum Science and Technology, Research School of Physics, The Australian National University, Canberra, ACT 2601, Australia}
\author{Alexander Yakimenko}
\affiliation{Department of Physics, Taras Shevchenko National University of Kyiv, 64/13, Volodymyrska Street, Kyiv 01601, Ukraine}
\altaffiliation[Also at ]{Dipartimento di Fisica e Astronomia 'Galileo Galilei' and Padua QTech Center, Università di Padova, via Marzolo 8, 35131 Padova, Italy}
\author{Eliezer Estrecho}%
\affiliation{Department of Quantum Science and Technology, Research School of Physics, The Australian National University, Canberra, ACT 2601, Australia}
\email{eliezer.estrecho@anu.edu.au}

\date{\today}

\begin{abstract}
Topological defects are central to modern physics, from spintronics to photonics, due to their robustness and potential application in information processing. In this work, we discuss topological point defects that spontaneously emerge at the imaginary Fermi arcs (degeneracy lines) in momentum space of two-dimensional systems described by non-Hermitian effective Hamiltonians. In particular, we consider a generic non-Hermitian Dirac model and a phenomenological model describing hybrid light-matter quasiparticles - exciton polaritons hosted in an optical microcavity. In both cases, the eigenenergies of the system have both real and imaginary parts and form two distinct bands corresponding to two (pseudo-)spin states. We describe the trajectories of the point defects characterized by integer-valued topological winding (vorticity) analytically and show that the defects with opposite vorticity annihilate with each other in the fully gapped phases, but are protected from annihilation by the non-Hermitian spectral degeneracies (exceptional and hybrid points) in the gapless phases. We also suggest that the signatures of these defects can be experimentally measured in an exciton-polariton system.
\end{abstract}

\maketitle

\section{Introduction}
Skyrmions are topologically protected field configurations, originally proposed as solutions of a nonlinear mesonic field theory in nuclear physics \cite{skyrme1961,skyrme1962}, and often interpreted in lower-dimensional systems as point-like defects.  In recent years, they were investigated in the broad context of spinor fields, such that of photons with distinct polarization states \cite{shen2024,guo2020,donati2016} and electronic spins in solid-state systems \cite{gobel2021,everschor-sitte2018,kovalev2018}. These defects are topologically protected, i.e. they preserve the associated integer-valued topological winding (such as vorticity and skyrmion number), and therefore have potential applications in the lossless transmission of quantum information and memory in spintronics devices \cite{gobel2021,everschor-sitte2018,kovalev2018,nagaosa2013}. 

Different types of point defects, such as anti-skyrmions, merons (half-skyrmions), and spin vortices, have attracted a great amount of interest in various physical systems, including but not limited to magnetic systems \cite{bogdanov1989,bogdanov1994,rößler2006,binz2006,tewari2007,hertel2006,tretiakov2007,muhlbauer2009}, nematic liquid crystals \cite{wright1989,aranson2019,fukuda2011,fumeron2021,fumeron2023,gim2017}, photonic \cite{shen2024,guo2020,donati2016}, plasmonic \cite{tsesses2018,tsesses2019,du2019,davis2020,bai2020,wang2020skyr,zhang2021}, and hybrid photonic (polaritonic) systems  \cite{cilibrizzi2016,flayac2013,vishnevsky2013,krol2021}. The topological defects can appear in position space due to the Dzyaloshinsky-Moriya interaction in magnetic systems \cite{dzyaloshinsky1958,moriya1960,bogdanov1989,bogdanov1994,rößler2006,binz2006,tewari2007,hertel2006,tretiakov2007,muhlbauer2009}, or due to artificial gauge field arising from the photonic spin-orbit interaction and anisotropy in photonic, plasmonic, and polaritonic systems \cite{donati2016,tsesses2018,tsesses2019,du2019,davis2020,bai2020,wang2020skyr,zhang2021,cilibrizzi2016,flayac2013,vishnevsky2013,krol2021}. 

They can also be generated in momentum space through a non-zero Berry curvature \cite{guo2020} or using an optical bound state in the continuum \cite{zhen2014,doeleman2018,jin2019,wang2020,bai2021,wu2022,rao2025}. Recently, we have also suggested that anti-merons and spin vortices can spontaneously form in momentum space due to the dynamics of wavepackets in non-Hermitian systems with gain and loss \cite{hu2023}. However, so far, little is known about the dynamical behaviour of these defects, their topological protection, and a possibility to observe them experimentally. 

In general, open-dissipative systems described by non-Hermitian effective Hamiltonians attract a lot of interests because they exhibit novel topological properties due to the combination of gain and loss, such as new topological invariants \cite{gong2018,shen2018,kawabata2019,yao2018,leykam2017,zhang2020} and edge states \cite{yao2018,lee2019,xiong2018,yokomizo2019,kunst2018,zhang2020,borgnia2020,okuma2020}. Moreover, one such non-Hermitian system -- exciton polaritons arising from strong coupling between the electron-hole pairs (excitons) in semiconductors and cavity photons \cite{carusotto2013,deng2010,spencer2021,pickup2020,estrecho2016} represents an accessible experimental platform that lends itself to exploration of non-Hermitian physics. Exciton polaritons are dissipative as a result of their photonic component and decay releasing photons. By measuring the photon emission, one can directly measure the physical observables, such as real and imaginary parts of their eigenenergy spectrum,  non-Hermitian spectral degeneracies, and non-Hermitian topological invariants \cite{gao2015,su2021,krol2022}. Exciton-polariton system is also intrinsically spinor, with the two components of the polariton (pseudo-)spin linked to the circular polarization of the photon emission. This makes the system highly suited for experimental probing of the (pseudo-)spin textures and defects. In our previous work, we examined the defects arising from an imaginary effective field due to the polarization-dependent photonic losses in a exciton-polariton system \cite{hu2023}. In this work, we provide comprehensive analysis of these emergent defects and their dynamics, with the focus on potential experimental signatures of the relevant effects.

First, we characterize the emergent topological defects using a simple non-Hermitian Dirac model. This model allows us to describe the basic features of the formation and dynamics of the point defects in the momentum space of a two-band system that can undergo a transition from a gapless to a gapped phase in the real part of the eigenenergy spectrum. The defining feature of these defects is that they are associated with degeneracy lines (imaginary Fermi arcs) of the imaginary part of the eigenenergy spectrum, and their dynamics is fully deterministic and analytically tractable. This sets the non-Hermitian systems apart from Hermitian systems, where the topological point defects typically emerge in position space and this spontaneous process is associated with the development of instabilities and turbulence due to nonlinearity \cite{kolmogorov1968,reeves2013,han2018,panico2023}. 

In this work, the term "topological" refers to the quantized winding of the local pseudospin vector around point defects in momentum space, characterized by integer-valued invariants such as vorticity and skyrmion number. These quantities are very robust under smooth deformations of the pseudospin texture and thus serve as local topological indices. However,  our results do not rely on a globally defined topological invariant (e.g., Chern number) and do not imply a topological phase of matter in the conventional bulk-boundary sense.

We then consider a phenomenological model describing an experimentally relevant exciton-polariton system with polarization-dependent photonic losses realized in a liquid-crystal-based microcavity. We provide a comprehensive analysis of the emergent topological defects and show that their dynamics is distinctly different in the two phases characterized by either gapped or gapless real-part spectrum of the system's complex eigenenergies. In the latter, gapless phase, the topological defects of different vorticity are protected from mutual annihilation by a non-Hermitian spectral degeneracy (exceptional or hybrid point).



\section{Pseudospin Dynamics}

\subsection{Point Defects in a Spinor Field}

First, we revisit the concept of spinor fields and the topological defects. For a quantum state in a two-level system, $\psi=(\psi^+,\psi^-)$, we can write down a set of Stokes vectors (pseudospins) $\mathbf{S}=(S_x,S_y,S_z)$ as $S_j=\langle\psi|\sigma_j|\psi\rangle/\langle\psi|\psi\rangle$ where $\sigma_j$ denotes the $j$-th two-by-two Pauli matrix. One example is a photonic or a polaritonic system with polarization splitting arising from anisotropy, TE-TM splitting, or Zeeman splitting \cite{kavokin2007}. In the basis of circularly-polarized mode where $\psi^+$ ($\psi^-$) corresponds to right (left) circular polarization, the Stokes vectors in the photonic and polaritonic systems can be expressed as: \cite{bleu2018,gianfrate2020}
\begin{equation}
    \begin{split}
        S_x=\frac{2\text{Re}[(\psi^+)^*\psi^-]}{|\psi^+|^2+|\psi^-|^2}\\
        S_y=\frac{2\text{Im}[(\psi^+)^*\psi^-]}{|\psi^+|^2+|\psi^-|^2}\\
        S_z=\frac{|\psi^+|^2-|\psi^-|^2}{|\psi^+|^2+|\psi^-|^2}.
    \end{split}
\end{equation}

The point defects in this spinor system can be classified using a topological invariant known as the skyrmion number, which measures the winding of the pseudospin around the core of the defect \cite{kovalev2018,everschor-sitte2018,gobel2021,shen2024} \cite{gobel2021,everschor-sitte2018,kovalev2018,shen2024}:
\begin{equation}
    N_{sk}=\frac{1}{4\pi}\int\mathbf{S}\cdot(\partial_{x}\mathbf{S}\times\partial_{y}\mathbf{S}) d^2\mathbf{r},
\end{equation}
where the integral is taken over the area from the core to the edge of the defect, which measures the winding of the pseudospin around the core of the defect \cite{kovalev2018,everschor-sitte2018,gobel2021,shen2024}. The skyrmion number can also be re-written in terms of two other topological indices:
\begin{equation}
    N_{sk}=\nu\cdot p,
\end{equation}
where $\nu$ denotes the vorticity and $p$ denotes the polarity. The vorticity characterizes the winding of the in-plane Stokes vectors $[S_x,S_y]$ around the defect core
\begin{equation}
    \nu=\frac{1}{2\pi}\oint d\mathbf{r}\cdot\nabla_\mathbf{r}\Phi(\mathbf{r}),
\end{equation}
where $\Phi(\mathbf{r})=\frac{1}{2}\arctan(S_y(\mathbf{r})/S_x(\mathbf{r}))$ \cite{gobel2021,arjas2024} is the azimuthal angle of the Stokes vector, and a vortex (anti-vortex) has $\nu=1$ ($\nu=-1$). 

The polarity $p$ measures the rotation of the out-of-plane Stokes vector $S_z$ from the core ($\mathbf{r}_{core}$) to the edge ($\mathbf{r}_{edge}$) of the defects \cite{kovalev2018,everschor-sitte2018,gobel2021,zheng2017} and is defined as:
\begin{equation}
    p=\frac{1}{2}\Big(S_z(\mathbf{r}_{core})-S_z(\mathbf{r}_{edge})\Big).
\end{equation}

\begin{figure}[t]
    \centering
    \includegraphics[width=0.45\textwidth]{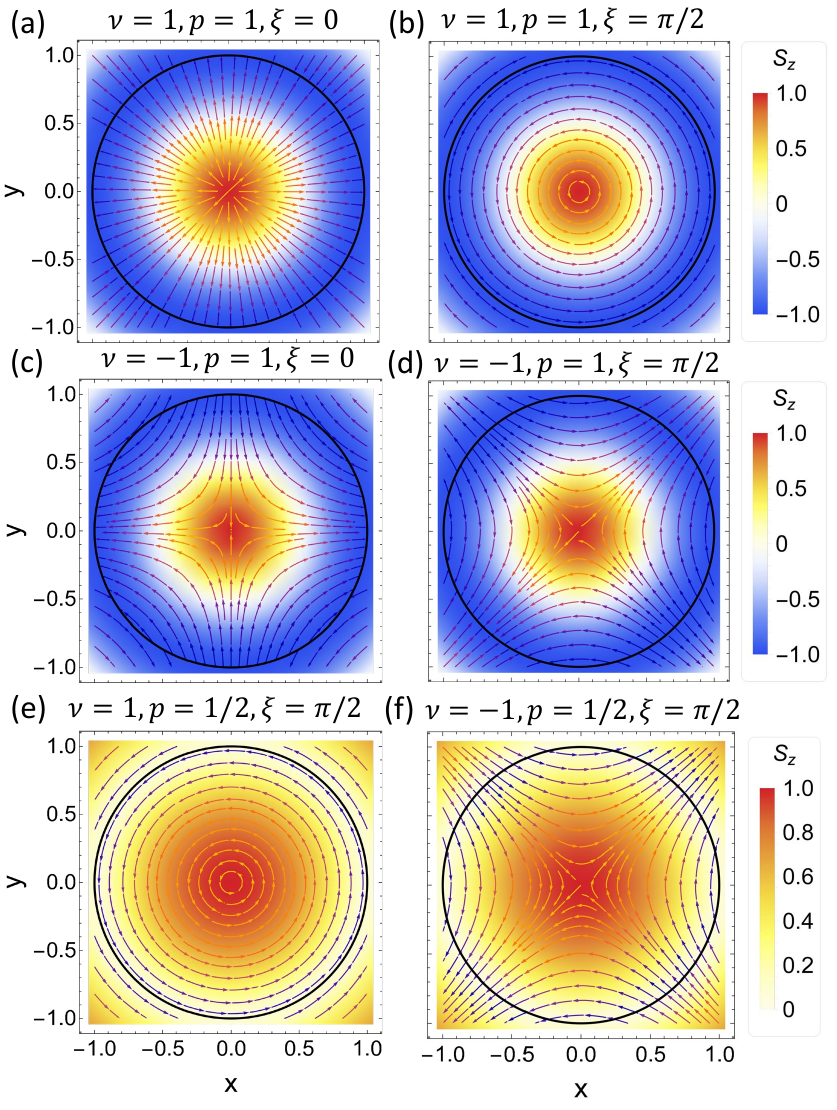}
    \caption{Spin textures of core-up defect of (a) a Neel-type (helicity $\xi=0$) and (b) a Bloch-type ($\xi=\pi/2$) skyrmions, (c,d) anti-skyrmions with helicities of (c) $\xi=0$ and (d) $\xi=\pi/2$, and (e, f) a meron and an anti-meron with helicity of $\xi=\pi/2$. The arrows represent $[S_x,S_y]$ and the color map represents $S_z$, the defects are centred at $\mathbf{r}_{core}=(0,0)$ and the black line outlines the edge of the defects, $\mathbf{r}_{edge}$ where $|\mathbf{r}_{edge}|=1$. Compared with these configurations, the core-down defects have the same in-plane spin textures but the opposite $S_z$.}
    \label{fig:0}
\end{figure}

In this work, we consider skyrmion numbers $N_{sk}=\pm1,\pm1/2$ corresponding to the common skyrmions and merons. Core-up (core-down) skyrmions have $S_z=1$ ($S_z=-1$) at the core with polarity of $p=1$ ($p=-1$) as $S_z(\mathbf{r}_{edge})\rightarrow-1$ ($S_z(\mathbf{r}_{edge})\rightarrow-1$), i.e. the core and edge have opposite spins. For a core-up defect, a skyrmion number of $N_{sk}=1$ ($N_{sk}=-1$) corresponds to a skyrmion (an anti-skyrmion) \cite{gobel2021,kovalev2018,everschor-sitte2018}. Figures~\ref{fig:0}(a-d) show examples of core-up skyrmions and anti-skyrmions. Conversely, for a core-down defect, we follow the convention in Refs.~\cite{kovalev2018,guo2020} and flip the sign so that a core-down skyrmion and a core-down anti-skyrmion is characterized by $N_{sk}=-1$ and $N_{sk}=1$, respectively. In this convention, a skyrmion (an anti-skyrmion) corresponds to a vortex (an anti-vortex) with an integer-valued polarity.

The polarity can also take half-integer values $p=\pm1/2$ corresponding to $S_z(\mathbf{r}_{edge})\rightarrow0$, i.e. the defect edge has no $S_z$-component. These half-skyrmions are called merons and anti-merons. Core-up merons and anti-merons have the skyrmion numbers of $N_{sk}=1/2$ and $N_{sk}=-1/2$, respectively \cite{gobel2021,kovalev2018,everschor-sitte2018}. Examples of core-up meron and anti-meron are shown in Figs.~\ref{fig:0}(e, f).

There are two types of common skyrmion textures distinguished by helicity $\xi$. In a polar coordinate system $(r,\phi)$ centred at the defect core ($r=0$ at $\mathbf{r}_{core}$), $\xi$ is defined by the global rotation of the in-plane Stokes vector with respect to the azimuthal coordinate $\phi$, $\Phi=\nu\phi+\xi$ \cite{gobel2021,kovalev2018,zhang2021}. The Neel-type skyrmion has zero azimuthal component of Stokes vectors, $S_\phi=0$, and a helicity of $\xi=0$, while the Bloch-type skyrmion has no radial component of Stokes vector $S_r=0$ and a helicity of $\pi/2$. Similarly, in Fig.~\ref{fig:0}(c,d), we show two anti-skyrmions with helicities of $\xi=0$ and $\xi=\pi/2$, respectively. In Fig.~\ref{fig:0}(e,f), we show a meron and an anti-meron, both with helicity of $\xi=\pi/2$. 

Although the skyrmion number and the associated quantities are usually defined in position space, they can also be defined in momentum space~\cite{guo2020}. For example, optical bound states in the continuum have been shown to occur at the centre of momentum-space polarization vortices  characterized by integer-valued vorticity \cite{zhen2014,doeleman2018,arjas2024}. In what follows, we discuss the emergence and dynamics of the pseudospin defects in momentum space of two-band non-Hermitian systems, including an experimentally relevant non-Hermitian model of photons and polaritons in a planar optical microcavity.

\subsection{Non-Hermitian Dirac Model}
The first model we consider is a non-Hermitian generalization of the well-known two-dimensional Dirac model, which is applicable to a wide range of physical systems, from describing the conical diffraction of circularly-polarized optical beams \cite{hamilton1837,lloyd1837} to the band structure of graphene\cite{armitage2018,castro2009,novoselov2004,novoselov2005}. The momentum-space Hamiltonian can be written in the form \cite{solnyshkov2021}:
\begin{equation}
    H=k_x\sigma_x+(k_y-i\kappa)\sigma_y+\Delta\sigma_z
\end{equation}
where $(k_x,k_y)$ are the components of the momentum, $\sigma_i$ are the $2\times2$ Pauli matrices, $\Delta$ describes a Zeeman splitting, and the term $i\kappa$ accounts for an imaginary effective magnetic field. Here, we set the constants $\hbar=1$ and $c=1$ for simplicity. 

\begin{figure*}[t]
    \centering
    \includegraphics[width=0.9\textwidth]{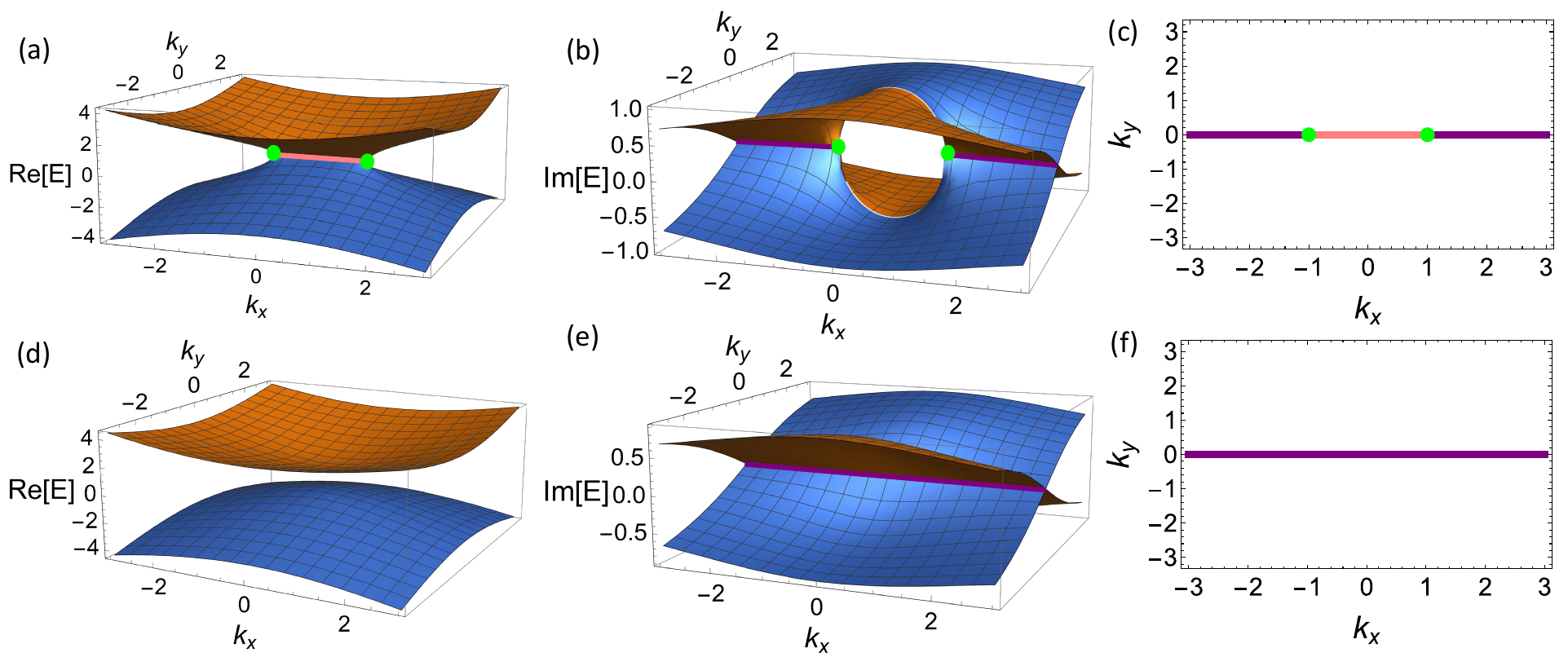}
    \caption{The (a) real and (b) imaginary parts of the eigenenergy of the non-Hermitian Dirac model ($\kappa=1$) in the gapless phase with $\Delta=0$ and the (d) real and (e) imaginary parts of eigenenergy  gapped phase with $\Delta=1.5$. The exceptional points (green dots), the bulk Fermi arc (pink line) and the imaginary Fermi arcs (purple lines) in the (c) gapless and the (f) gapped phase. The exceptional points and Fermi arcs in (a,b,e) are a guide to the eye.} 
    \label{fig:1}
\end{figure*}

This model has the complex eigenenergy  $E_\pm=\pm E$, where  $E=\sqrt{k_x^2+(k_y-i\kappa)^2+\Delta^2}$. The two eigenenergy bands in momentum space, shown in Fig. \ref{fig:1}, have two spectral degeneracies (exceptional points) at $\mathbf{k}_{EP}=[\pm\sqrt{\kappa^2-\Delta^2},0]$, when $|\Delta|<|\kappa|$. These are singular points (where the two eigenstates coalesce) that carry non-zero spectral windings \cite{shen2018}. The exceptional points are connected by the bulk Fermi arc (degeneracy line) along the $k_x$-axis at $|k_x|<|k_{EP,x}|$, where the real parts of the bands cross, i.e. $Re[E]=0$, and by the imaginary Fermi arc at $|k_x|>|k_{EP,x}|$ where the imaginary parts of the bands cross, i.e. $Im[E]=0$, as seen in Fig. \ref{fig:1}(c). 

When $|\Delta|=|\kappa|$, the two exceptional points merge into a single hybrid point, which is a singular point with zero spectral winding \cite{shen2018}, at $k_x=0$ and the bulk Fermi arc vanishes. When $|\Delta|>|\kappa|$, the degeneracy is lifted and a full gap opens in the real part of the eigenenergies, while the imaginary Fermi arc remains and extends to the entire $k_x$-axis. The annihilation of EPs and the opening of the gap in this model signifies a topological phase transition, since the band structure would lose the non-zero spectral winding around each EP, while the gap carries a non-zero Chern number.

\begin{figure}[t]
    \centering
    \includegraphics[width=0.45\textwidth]{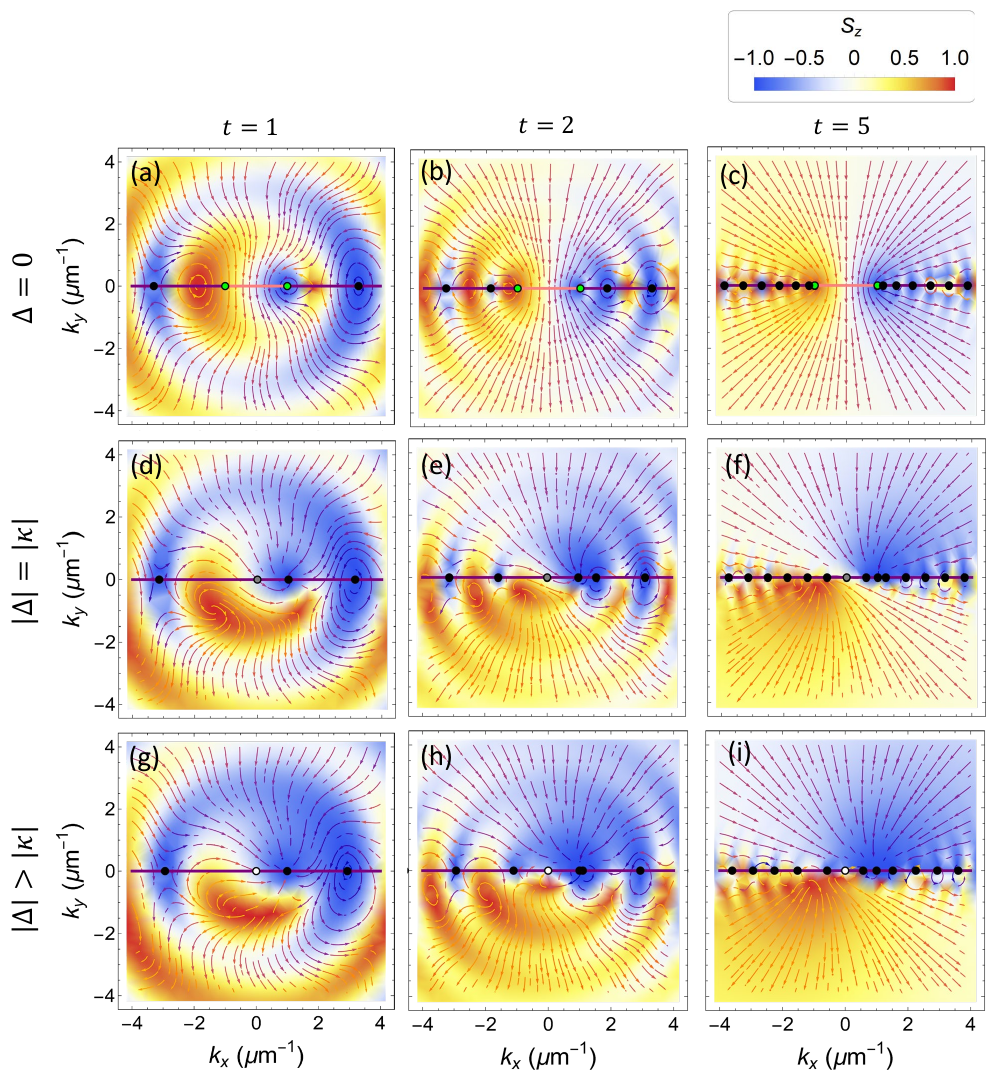}
    \caption{Snapshots of the defects arising in the non-Hermitian Dirac model ($\kappa=1$) in (a-c) a gapless phase with two exceptional points (green dots) for $\Delta=0$, (d-f) a gapless phase with a hybrid point (gray dots) for $\Delta=1$ and (g-i) a gapped phase for $\Delta=1.5$. 
    The initial condition is a spin-down state. Only the core-down defects are highlighted by the black dots for simplicity. The core-up defects migrate from the imaginary Fermi arc as $\Delta$ increases. The bulk and imaginary Fermi arcs are denoted by the pink and the purple lines, respectively. The points where $\min(\text{Re}[E])$ are marked by the white dots.} 
    \label{fig:2}
\end{figure}

As previously discussed in Ref.~\cite{hu2023}, the dynamics of a wavepacket in the system governed by the non-Hermitian Hamiltonian, leads to the formation of pseudospin defects on the imaginary Fermi arcs. Example snapshots are shown in Fig.~\ref{fig:2}, showing staggering core-up and core-down defects being generated along the imaginary Fermi arcs in the gapless phase at $\Delta=0$ [Figs.~\ref{fig:2}(a-c)]. At $\Delta=0$, the core-up (core-down) defects along the positive (negative) $k_x$-axis have vorticity of $\nu=-1$ and polarity of $p=1/2$ ($p=-1/2$), showing them to be anti-merons. On the other hand, the core-down (core-up) defects along the positive (negative) $k_x$-axis have vorticity of $\nu=1$ but they do not have a well-defined edge and do not have integer nor half-integer valued polarity. We refer to them as pseudospin vortices. In Figs.~\ref{fig:2}(d-i), as $\Delta$ increases, the exceptional points annihilate, and the gap opens, the core-down defects stay on the imaginary Fermi arcs, while the core-up defects are pushed off the arcs. At the same time, the pseudospin textures gain $S_z=\pm1$ on the two sides of the imaginary Fermi arcs. Subsequently, the anti-merons no longer have polarity of $p=\pm1/2$, and become pseudospin anti-vortices.

We note that the following analytical results are derived under the assumption of a spin-polarized initial condition, which simplifies the pseudospin evolution and allows for clear identification of the defect structure. The evolution of more general initial states, including mixed or spatially modulated pseudospin textures, remains an open question and may lead to richer defect dynamics. In the case where the initial wavefunction of the system $\psi(0)$ is in a spin-up (spin-down) configuration, we can show analytically that the centres of the emergent core-up (core-down) pseudospin point defects are located at $\mathbf{k}^*(t)=(\pm k_{x,n}^*(t),0)$ where
\begin{equation}\label{eq: dirac defect}
    k_{x,n}^*=\sqrt{k_{EP,x}^2+E_n^2(t)}.
\end{equation}
Here, $*$ indicates the momentum space coordinates of the defects and $E_n(t)$ satisfies the quantization condition $E_n(t)=n\pi\hbar/t$ and $n\in\mathbb{Z}$ is an integer index of the defects. The evolution of the pseudospin follows $\psi(t)=\hat{T}\psi(0)$, where $\hat{T}=\exp(-{iHt}/\hbar)$, which is linear and deterministic. The quantization condition corresponds to the moments of time when the time evolution operator $\hat{T}$, given by
\begin{equation}
    \hat{T}=\begin{pmatrix}
        \cos{\frac{Et}{\hbar}}-\frac{i\Delta}{E}\sin{\frac{Et}{\hbar}} && -i\frac{k_x-i(k_y-i\kappa)}{E}\sin{\frac{Et}{\hbar}}\\
        -i\frac{k_x+i(k_y-i\kappa)}{E}\sin{\frac{Et}{\hbar}} &&  \cos{\frac{Et}{\hbar}}+\frac{i\Delta}{E}\sin{\frac{Et}{\hbar}}
    \end{pmatrix}
\end{equation}
is the identity matrix, i.e. $\hat{T}$ maps $S_z=\pm1$ to itself.

Similarly, when $\hat{T}$ is an anti-diagonal matrix, it maps $S_z=\pm1$ to $S_z=\mp1$ and generates the defects with the core orientation opposite to the initial condition. At $\Delta=0$, this condition is satisfied along the imaginary Fermi arcs at $\mathbf{k}^*(t)=(\pm k_{x,m}^*(t),0)$ with $k_{x,m}^*$ given by Eq. (\ref{eq: dirac defect}) and the 
quantization condition $E_m=((2m+1)/2)\pi\hbar/t$, where $m$ is an integer. 

When $\Delta\neq 0$, for the initial spin-up (spin-down) configuration, only the defects with $S_z=1$ ($S_z=-1$) will stay on the imaginary Fermi arcs with their location in momentum space $\mathbf{k}^*$ determined by Eq.~(\ref{eq: dirac defect}), while the defects with $S_z=-1$ ($S_z=1$) will be pushed away from these arcs. This is because, when $\Delta\neq 0$, the condition for $\hat{T}$ to be anti-diagonal is satisfied for $E=\pm i\Delta\tan{Et/\hbar}$. The position of the defects $\mathbf{k}^*$ needs to be determined numerically and may no longer be on the imaginary Fermi arcs. In the rest of this work, we consider the initial spin-down configuration and focus our analysis on the dynamics of the core-down ($S_z=-1$) defects, since only these defects can be described analytically for all values of $\Delta$. 

Note that in the fully gapped phase, the pair of pseudospin defects generated at opposite momenta $\mathbf{k}^*=(\pm k_{x,n}^*(t),0)$ with the same index $n$ have the same polarity $p$, but opposite vorticity $\nu$, therefore they also have the opposite skyrmion numbers. The neighboring defects at $\mathbf{k}^*_n$ and $\mathbf{k}^*_{n+1}$ also have opposite vorticity $\nu$, but they have opposite polarity $p$, and therefore the same skyrmion numbers. The typical pseudospin textures associated with such defects and their time evolution is presented in Fig. \ref{fig:2} for the three different phases: gapless, with two exceptional points; gapless with a hybrid point at the origin, and fully gapped, with no degeneracy in the real part of the eigenenergy. 

\begin{figure}[t]
    \centering
    \includegraphics[width=0.5\textwidth]{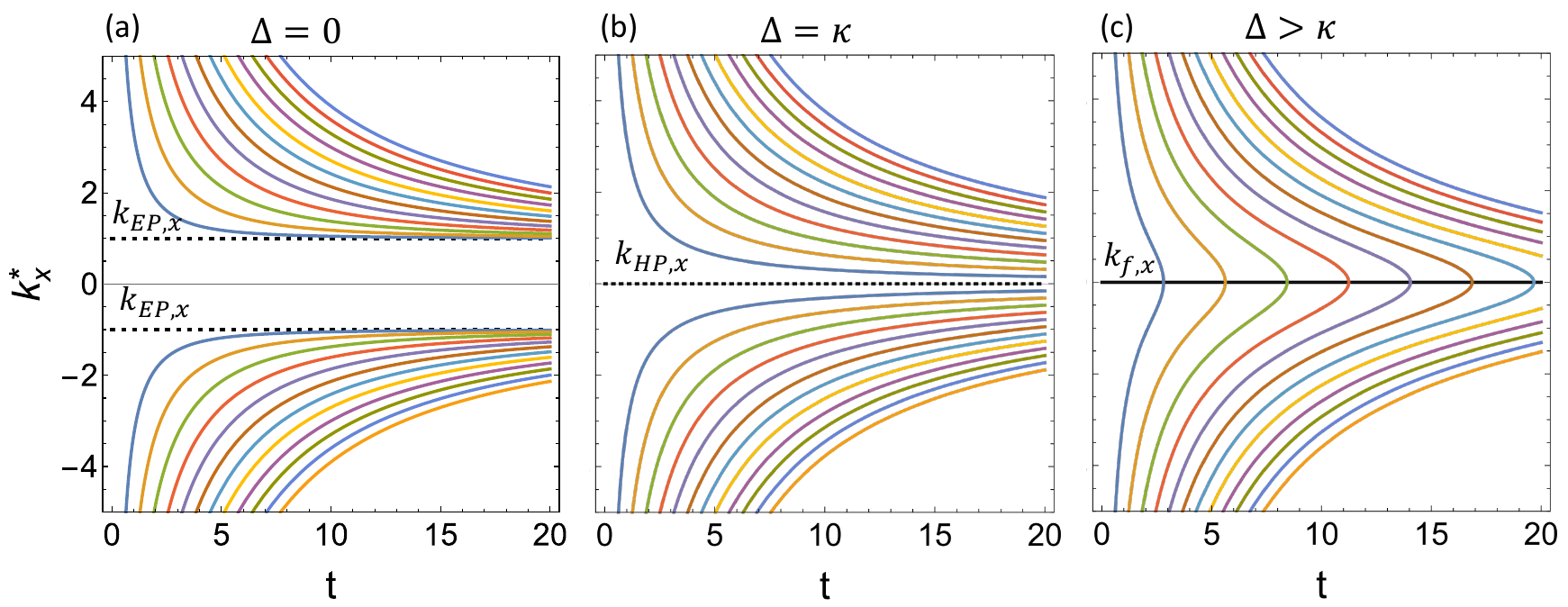}
    \caption{Position of the core-down defects on the imaginary Fermi arcs as a function of time in (a) a gapless phase with two exceptional points for $\Delta=0$, (b) another gapless phase with a hybrid point for $\Delta=1$ and (c) a fully gapped phase for $\Delta=1.5$. Different colors correspond to different $n$ up to $n=12$. Positions of the exceptional points ($k_{EP,x}$) in (a) and the hybrid point ($k_{HP,x}$) in (b) are marked with horizontal dashed lines, and $min(\text{Re}[E])$ is marked with a solid line ($k_f,x$).} 
    \label{fig:3}
\end{figure}

In the fully gapped phase, $|\Delta|>|\kappa|$, the two defects with opposite skyrmion numbers propagate along the imaginary Fermi arc until they annihilate with each other at the time $t=n\pi\hbar/\sqrt{\Delta^2-\kappa^2}$ at $k_x=0$, as shown in Fig. \ref{fig:3}(c). However, in the gapless phase with $|\Delta|\leq|\kappa|$, the defects cannot pass through the degeneracy so the defects accumulate at $\mathbf{k}^*\rightarrow\mathbf{k}_{EP}$ as $t\rightarrow\infty$, as seen in Figs. \ref{fig:3}(a,b). As the pairs of defects with opposite skyrmion numbers cannot move past the degeneracy, they do not annihilate, and the number of pseudospin defects will accumulate. With time, the defects become increasingly densely packed and move closer to the degeneracy, but never dissipate nor annihilate. Thus, the non-Hermitian spectral degeneracies, i.e. the hybrid and exceptional points provide topological protection to the pseudospin defects on the imaginary Fermi arcs. 

Note that despite the accumulation and increasing density of the defects with time, the total (zero) skyrmion number in the system remains unchanged because the defects with opposite skyrmion numbers are always generated in pairs.  

We note that, since the imaginary Fermi arc in the non-Hermitian Dirac model extends to $k_x=\pm\infty$, the defects are generated at $k_x^*\rightarrow\infty$ at $t\rightarrow0$ (see Fig. \ref{fig:3}). Therefore, this simple model cannot provide insights into the early-time dynamics of the defects. In what follows, we consider a more realistic and experimentally relevant non-Hermitian model, whose Fermi arcs are confined within finite regions in the momentum space, enabling richer defect dynamics.

\begin{figure*}[ht]
    \centering
    \includegraphics[width=0.9\textwidth]{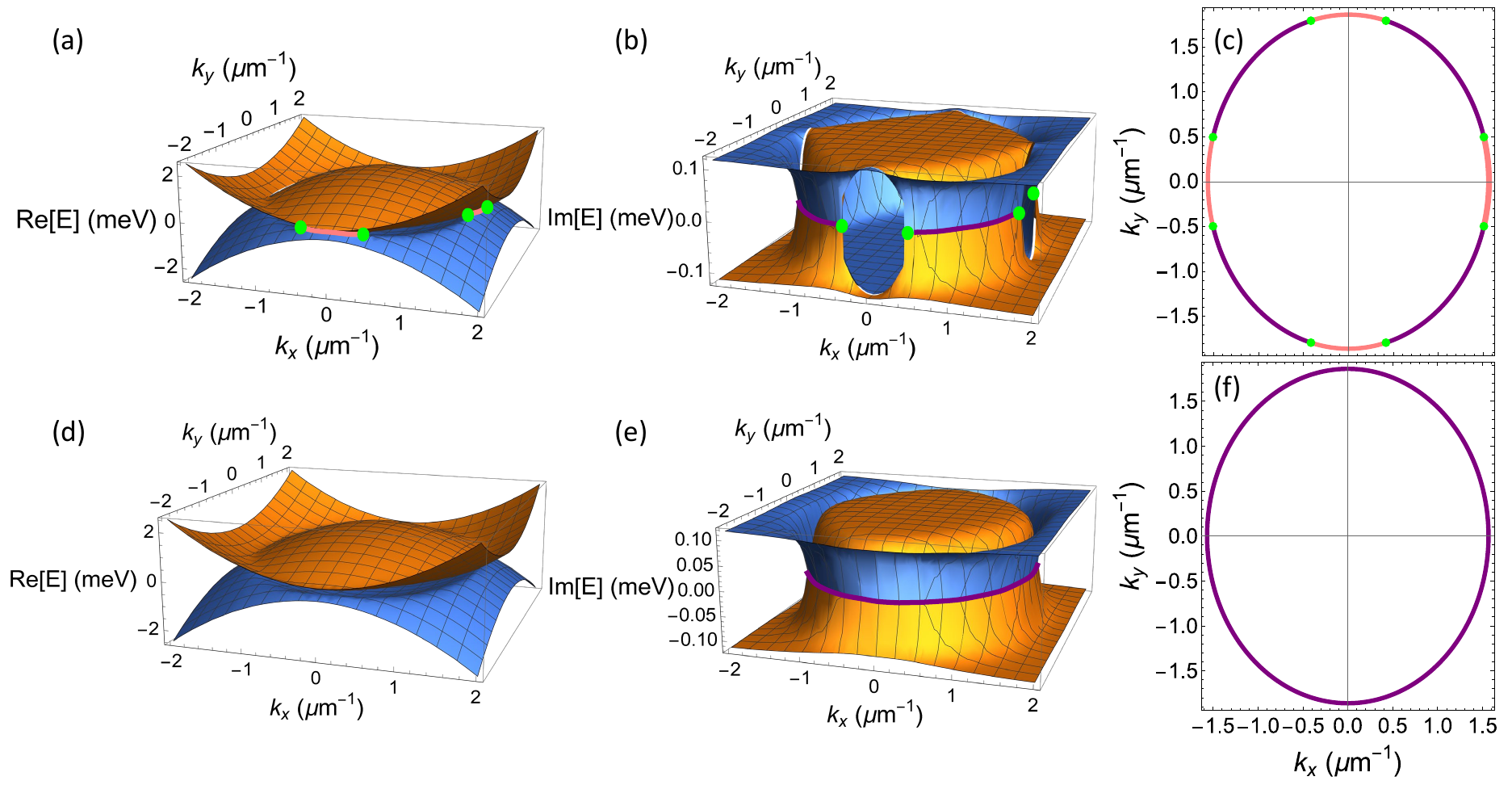}
    \caption{The (a) real and (b) imaginary parts of the mean-subtracted eigenenergy $E$ of the LC microcavity exciton polaritons in the gapless phase with $\Delta=0$ and the (d) real and (e) imaginary parts of $E$ in the gapped phase with $\Delta=1.5\times10^{-4} eV$. (c) Momentum-space structure of the spectral degeneracies showing the exceptional points (green dots), the bulk Fermi arcs (pink arcs) and the imaginary Fermi arcs (purple arcs) for (c) $\Delta=0$ eV and (f) $\Delta=1.5\times10^{-4} eV$.} 
    \label{fig:4}
\end{figure*}

\subsection{Non-Hermitian Exciton-Polariton Model}
The second model we consider in this work is a two-band non-Hermitian system with spin-orbit coupling. These features are experimentally realized in a liquid-crystal (LC) based anisotropic optical microcavity supporting the formation of exciton polaritons \cite{lempicka2022electrically,polimeno2021tuning,krol2022}.. The system is described by the effective non-Hermitian Hamiltonian \cite{krol2022}
\begin{equation}
\begin{split}
    H=&H_0+\mathbf{H}\cdot\boldsymbol{\sigma},\\
    \mathbf{H}=&(H_x,-2\beta k_xk_y,\Delta),
    \end{split}
\end{equation}
where $\boldsymbol{\sigma}=(\sigma_x,\sigma_y,\sigma_z)$ is the Pauli vector, and the components of the Hamiltonian are defined as:
\begin{equation}
\begin{split}
    H_0=&E_0-i\Gamma_0+\frac{\hbar^2k_x^2}{2m_x}+\frac{\hbar^2k_y^2}{2m_y},\\
    H_x=&\alpha-\beta(k_x^2-k_y^2)-\beta'(k_x^2+k_y^2)+i\delta\Gamma,\\
\end{split}
\end{equation}
Here, $\alpha$ is the cavity anisotropy that splits the linearly-polarized photonic modes, $\beta$ is the the photonic spin-orbit coupling arising from TE-TM splitting of the cavity modes, $\beta'$ represents the enhancement in the TE-TM splitting arising from the orientation of the LC molecules, $\Delta$ represents a Zeeman splitting arising from an external magnetic field which splits the energy of circularly-polarized modes, and $\delta\Gamma$ represents the splitting between the linewidths of the linearly-polarized modes due to polarization-dependent losses in the system \cite{krol2022}.

The system has the eigenenergies $E_\pm=H_0\pm E$, where $E=\sqrt{H_x^2+4\beta^2 k_x^2k_y^2+\Delta^2}$ denotes the mean-subtracted eigenenergy, which determines the pseudospin dynamics. In the Hermitian limit, there are four Dirac points when $\beta'>\beta$, but the non-Hermiticity splits each Dirac point into a pair of exceptional points \cite{krol2022} (see Fig. \ref{fig:4}). A non-zero Zeeman field with $|\Delta|>|\delta\Gamma|$ annihilates the exceptional points and opens a real gap in the band structure. 

The bulk and imaginary Fermi arcs and the exceptional points are situated at locations in momentum space where the real and imaginary effective magnetic fields becomes perpendicular to each other. This condition is satisfied at
\begin{equation}
    k_F^2(\phi)=\frac{\alpha}{\beta'+\beta\cos{2\phi}}
\end{equation}
which is an ellipse that describes the degeneracy and the Fermi arcs,where $\phi$ is the polar angle \cite{krol2022} [Fig.~\ref{fig:4}(c)]. The band structure is gapless when $|\Delta|<|\delta\Gamma|$, and has eight exceptional points at $\mathbf{k}_{EP}=(k_F(\phi_{EP}),\phi_{EP})$ (see Fig. \ref{fig:4}(c)), with the analytical expression for $\phi^{EP}$ as a function of the system parameters presented in the Appendix, Eq. (\ref{phiep}). 

When $|\Delta|=|\delta\Gamma|$, each pair of exceptional points merges into a hybrid point at $\phi=m\pi/2$, $m=0,1,2,3$, and at a larger $|\Delta|>|\delta\Gamma|$ a gap opens in the real parts of the eigenenergies. Interestingly, this point where the exceptional points and the hybrid point are annihilated also corresponds to $\min(\text{Re}[E])$.

Similarly to the non-Hermitian Dirac model, the pseudospin defects emerge on the imaginary Fermi arcs at $\mathbf{k}^*=(k_F(\phi^*),\phi^*)$ where the exact functional dependence of $\phi^*$ on the parameters of the model is given in the Appendix A, Eq. (\ref{phistar}).

In contrast with the non-Hermitian Dirac model, where the defects emerge at infinity, the imaginary Fermi arcs in the exciton-polariton system have a finite length and the defects emerge at finite values of $\mathbf{k}$. The pair of defects are generated at $\mathbf{k}^*=(k_F(\phi_{i}),\phi_{i})$ periodically at $t=nT_i$, where
\begin{equation}
    \begin{split}
        T_i=&\pi\hbar\sqrt{\frac{\beta'^2-\beta^2}{\alpha^2\beta^2+(\beta'^2-\beta^2)(\Delta^2-\delta\Gamma^2)}}\\
        \phi_i=&\pm\arccos{\bigg( \pm\sqrt{\frac{\beta'-\beta}{2\beta'}} \bigg)}.
    \end{split}
\end{equation}
The index $n$ in $T_i$ corresponds to the same index in $\phi^*$ (\ref{phistar}) and the two $\pm$-signs in $\phi_i$ are independent, giving four values of $\phi_i$. 

In the fully gapped phase, a pair of defects with the same index $n$ and polarity $p$, but opposite vorticity $\nu$ will emerge at the same $\phi_i$, then each defect will propagate along the imaginary Fermi arcs towards $\mathbf{k}*=(k_F(m\pi/2),m\pi/2)$, where $m=0,1,2,3$, and annihilate with another defect carrying the opposite skyrmion number at $t=nT_f$ where
\begin{equation}
    T_f=\frac{\pi\hbar}{\sqrt{\Delta^2-\delta\Gamma^2}}.
\end{equation}
Note that the periods for creation $T_i$ and annihilation $T_f$ are different, so the defects will still accumulate.

 In the gapless phase with either exceptional points or hybrid points, similarly to the non-Hermitian Dirac model, the defects with opposite skyrmion number will propagate along the imaginary Fermi arcs towards the non-Hermitian degeneracies, which protect them from annihilation. Notably, along the imaginary Fermi arcs, the real part of the eigenenergy $\text{Re}[E]$ reaches its maxima exactly at $\phi=\phi_i$, and its minima at $\phi=m\pi/2$ in the fully gapped phase, or at the degeneracy in the gapless phase. 

\begin{figure}[t]
    \centering
    \includegraphics[width=0.5\textwidth]{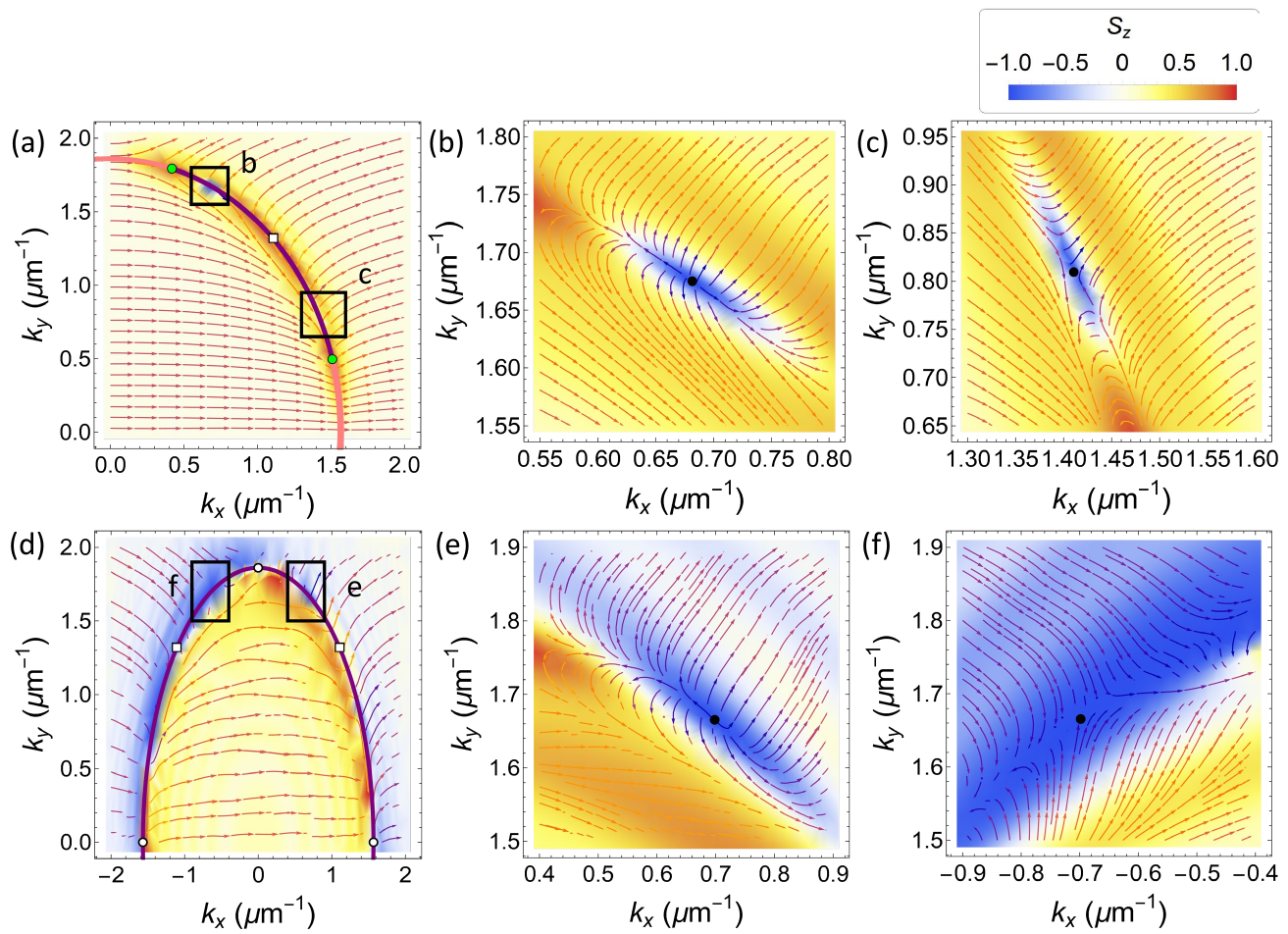}
    \caption{(a) A pair of point defects with the opposite vorticity, $\nu$, in the gapless phase with $\Delta=0$ eV, at $t=15$ ps . (b,c) The close-up on the two defects with (b) $\nu=1$ and  (c) $\nu=-1$ shown in (a). (d) A pair of point defects  with the opposite vorticity, $\nu$, at in the fully gapped phase with $\Delta=15\times10^{-4}$ eV, at $t=10$ ps. (e,f) The close-up of the two defects with (e) $\nu=1$ and (f) $\nu=-1$ shown in (d) that propagate towards $\phi=\pi/2$ before annihilating. The white square with black outline in (a,d) correspond to $\max(\text{Re}[E])$ where the defects emerge from. The green dots with black outline in (a) and the white dots with black outline in (d) correspond to the $\min(\text{Re}[E])$, where the defects propagate towards, which also correspond to the degeneracies in the gapless phase.} 
    \label{fig:5}
\end{figure}

\begin{figure}[ht]
    \centering
    \includegraphics[width=0.5\textwidth]{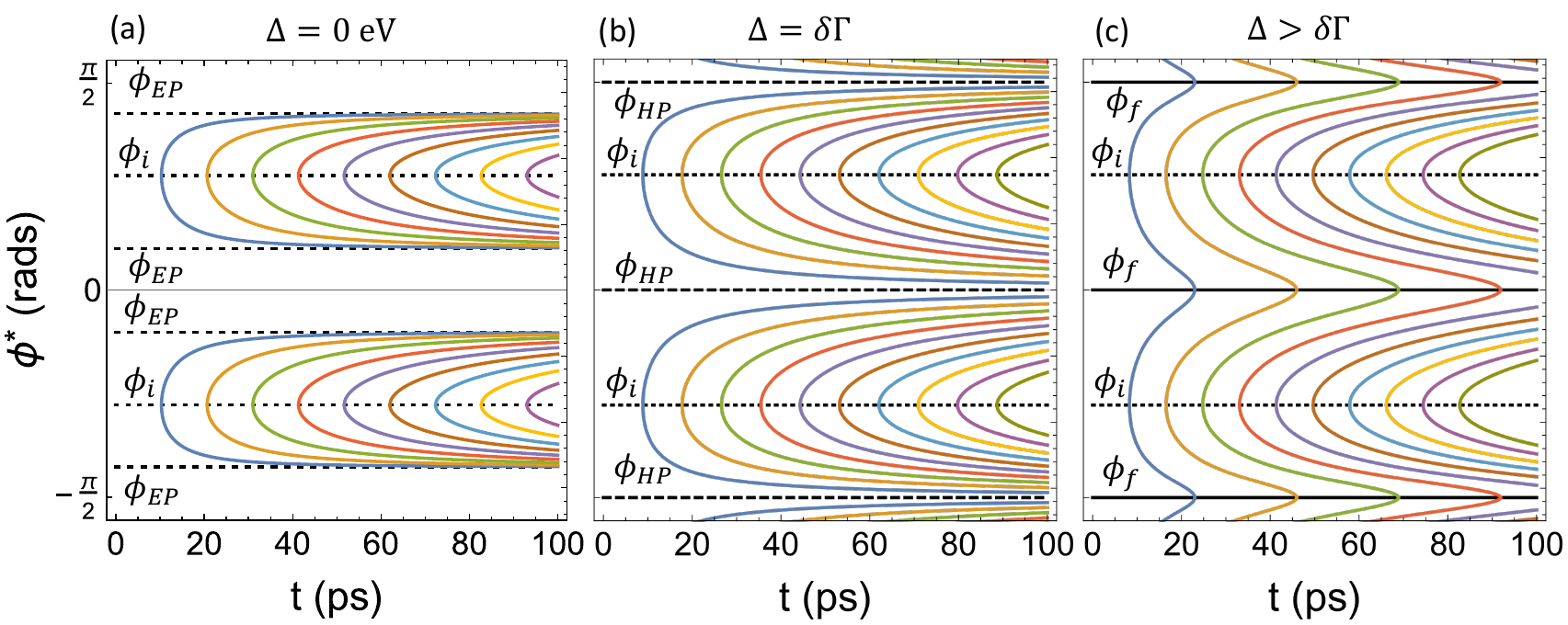}
    \caption{Position of the core-down defects on the imaginary Fermi arcs plotted with time in (a) a gapless phase with exceptional points (denoted as $\phi_{EP}$) for $\Delta=0$ eV, (b) another gapless phase with hybrid points (denoted as $\phi_{HP}$) for $\Delta=\delta\Gamma=1.2\times10^{-4}$ eV and (c) a fully gapped phase for $\Delta=1.5\times10^{-4}$ eV, where the defects are annihilated at $\phi_f$. Here, different colors correspond to different $n$.} 
    \label{fig:6}
\end{figure}

\begin{figure}[t]
    \centering
    \includegraphics[width=0.4\textwidth]{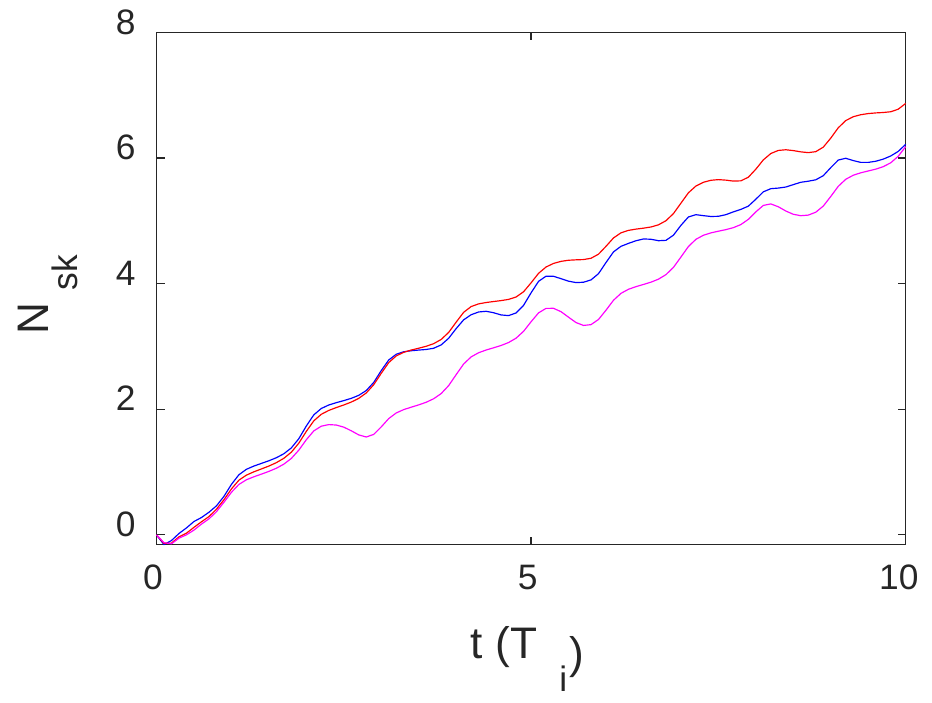}
    \caption{The skyrmion number calculated in the region between $k\in[0,2]$ and $\phi\in[0,\arccos\bigg({\sqrt{\frac{\beta'-\beta}{2\beta'}}\bigg)\approx0.87}]$ for $\Delta=0$ (blue line), $\Delta=\delta\Gamma$ (red line) and $\Delta>\delta\Gamma$ (purple line). Here, $t$ is in the unit of $T_i$, which varies with $\Delta$.} 
    \label{fig:7}
\end{figure}

The qualitative change in the behavior of the pseudospin defects in the gapless and the gapped phase is also reflected in the skyrmion number in momentum space. In Fig.~\ref{fig:7}, we plot the skyrmion number calculated within the region in momentum space between $\phi=\arccos{\sqrt{(\beta'-\beta)/(2\beta')}}$, which is one of the $\phi_i$ where the defects are generated, and $\phi=0$, where the defects propagate towards. We chose this region to account for only one of the defects in each of the emerging (annihilating) pairs with the opposite skyrmion numbers. 
In the gapless phases, the total skyrmion number in this momentum-space region increases monotonically [red line in Fig.~\ref{fig:7}] as a result of more and more defects accumulating near the spectral degeneracy. However, once a gap is opened, we can see periodic dips in the skyrmion number [blue and purple lines in Fig.~\ref{fig:7}] that correspond to the annihilation events. The dips become more pronounced as the gap increases.

\begin{figure}[h]
    \centering
    \includegraphics[width=0.45\textwidth]{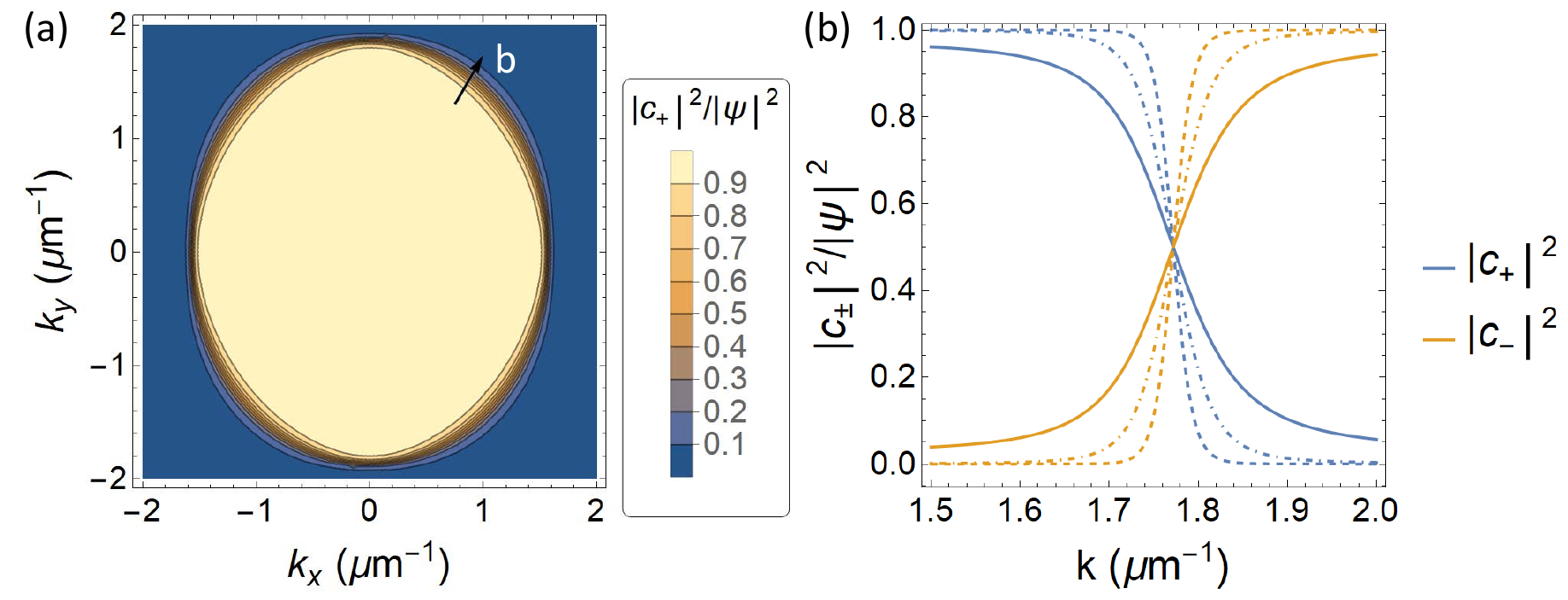}
    \caption{(a) The normalized overlap with the upper eigenstate at $t=5$ ps. (b) The cross section of normalized $|c_\pm|^2$ along the $k$-direction at $\phi=\pi/3$ marked with the black arrow in (a) at $t=5$ ps (solid lines), $t=10$ ps (dotted dashed lines) and $t=20$ ps (dashed lines).} 
    \label{fig:8}
\end{figure}

Interestingly, the imaginary Fermi arcs in this system play a role similar to that of domain walls \cite{takeuchi2022,kokubo2021}, but instead of spin-up and spin-down regions, they separate regions that align with the upper and the lower eigenstates [Fig. \ref{fig:8}]. For an initial spin-down configuration, the overlap between the wavefunction and the upper (lower) eigenstates can be quantified as [see Appendix B for derivation]:
\begin{equation}
    \frac{|c_\pm|^2}{|\psi|^2}=\frac{e^{\pm\frac{2\text{Im}[E]t}{\hbar}}}{2\cosh{\frac{2\text{Im}[E]t}{\hbar}}}
\end{equation}
where $c_\pm$ denotes the components of the upper and lower eigenstates in the wavefunction $|\psi\rangle=c_+|\psi_+^R\rangle+c_-^2|\psi_-^R\rangle$ and $|\psi|^2=|c_+|^2+|c_-|^2$. In the case of spin-domain walls, skyrmions can be generated as a result of Kelvin-Helmholtz instability, as discussed in Refs. \cite{kokubo2021,takeuchi2022,huh2024}. However, in contrast with the quantum hydrodynamics where the domain wall emerges due to the nonlinear interactions, in the linear system considered here the pseudo-spin domains and point defects emerge due to the non-Hermiticity and the resulting complex structure of the momentum-space eigenenergies. The possibility that this non-Hermiticity may lead to effectively nonlinear dynamics of spin and pseudospin \cite{graefe2008,graefe2010,graefe2010iop,pi2024,cius2022,zheng2021,wu2021} requires further investigation.

\section{Conclusion}
In conclusion, we describe the process of generation, annihilation, and dynamics of emergent momentum-space pseudospin defects in a non-Hermitian two-band system. The spectrum exhibits a transition between a fully gapless and a gapped phase, the latter characterized by the presence of non-Hermitian degeneracies (exceptional and hybrid points). In particular, we consider a generic non-Hermitian Dirac model, and a model describing a hybrid photonic (exciton polariton) system with optical spin-orbit coupling and polarization-dependent losses. Using the simple case where the wavefunction is initially in a circularly-polarized state, we show analytically that the defects are generated, in pairs, at the maxima of mean-subtracted eigenenergy, then propagate towards the minima of the mean-subtracted eigenenergy. We also show that the defects with opposite vorticity will be annihilated if there is a real gap in the spectrum, but in the gapped phases, they will be protected by the non-Hermitian degeneracy. This process preserves the total (zero) skyrmion number in the system. In contrast with quantum hydrodynamics of turbulent systems, the location and trajectories of the skyrmions in momentum space are fully deterministic and are determined by the location of eigenenergy degeneracy lines (imaginary Fermi arcs) in momentum space.

Since the pseudospin texture in an exciton-polariton system can be experimentally measured \cite{su2021,krol2021,bleu2018,gianfrate2020}, by measuring the pseudospin and the skyrmion number in a certain region of the momentum space at different times, one can potentially detect the signature of the transition between the gapped and the gapless phase.

We also show that an imaginary Fermi arc forms a domain wall separating the regions where the pseudospins align with the two different eigenstates of the non-Hermitian system. It is therefore tempting to draw parallels with previous studies connecting the generation of skyrmions in a Hermitian spinor system to the domain wall instabilities \cite{takeuchi2022,kokubo2021,huh2024}, but a closer link between these effects remains to be uncovered. The time-resolved defect dynamics can be observed using a streak camera, which takes a screen shot of the distribution and polarization textures of the polaritons every 2 to 4 ps, allowing us to directly observe the pseudospin dynamics ~\cite{estrecho2021,hu2023}. The nucleation of the pseudospin defects in momentum space will then give a clear signature of the imaginary Fermi arcs. Our study highlights the rich new physics of non-Hermitian systems and the role played by the imaginary Fermi arcs in the experimentally observable dynamical effects.


\begin{acknowledgments}
We acknowledge support from the Australian Research Council (ARC) through the Centre of Excellence Grant CE170100039 and the Discovery Early Career Researcher Award DE220100712, and Australian Government Research Training Program (RTP) Scholarship. This work was supported by the Australian Academy of Science Ukraine-Australia Research Fund (Grant No. STV00021R2).
\end{acknowledgments}

\begin{widetext}
\appendix
\section{Exceptional points and point defects in the exciton-polariton model}

The band structure of the non-Hermitian exciton-polariton model considered here is gapless when $|\Delta|<|\delta\Gamma|$, and has eight exceptional points. Their location in momentum space can be described by $\mathbf{k}=(k_F(\phi_{EP}),\phi_{EP})$ (with $k_F$ defined in the main text)  with the polar angle:
\begin{equation}\label{phiep}
    \phi_{EP}=\pm\arccos{\bigg(\pm\sqrt{\frac{\gamma\beta+\beta'(\Delta^2-\delta\Gamma^2)\pm\alpha\sqrt{\gamma\beta^2+\beta'^2(\Delta^2-\delta\Gamma^2)}}{2\beta\gamma}}\bigg)},
\end{equation}
where $\gamma=\alpha^2-(\Delta^2-\delta\Gamma^2)$ and the sign permutations $\pm$ are independent from each other.

The location of the pseudospin defects can also be defined by the polar angle $\phi^*$, where
\begin{equation}\label{phistar}
    \begin{split}
        \phi^*=&\frac{1}{2}\arctan{B_\pm/A_\pm}+m\pi\\
        A_\pm=&\frac{\beta'\delta E}{\beta(\gamma+E_n^2)}\pm\frac{\alpha\sqrt{\beta^2(\gamma+E_n^2)+\beta'^2\delta E}}{\beta(\gamma+E_n^2)}\\       
        B_\pm=&\pm\frac{\sqrt{\delta E}}{|\beta(\gamma+E_n^2)|}\sqrt{-2\alpha\beta'\sqrt{\beta^2(\gamma+E_n^2)+\beta'^2\delta E}-(\beta^2\gamma+\beta'^2\gamma')-E_n^2(\beta^2-\beta'^2)}.
    \end{split}
\end{equation}

Here $n\in\mathbb{N}$, $m\in\mathbb{Z}$, $\gamma'=\alpha^2+(\Delta^2-\delta\Gamma^2)$, $\delta E=\Delta^2-\delta\Gamma^2-E_n^2$, $E_n=n\pi\hbar/t$, and the sign permutations $\pm$ are independent.

\section{Imaginary Fermi arcs as Domain Walls}
In non-Hermitian systems, the coefficients $c_\pm$ can be calculated using biorthogonality as
\begin{equation}
    c_\pm=\langle\psi_\pm^L|\psi(t)\rangle.
\end{equation}
In a two-band non-Hermitian system, the left eigenstates have the general forms of
\begin{equation}
    \langle\psi^L_\pm|\propto\begin{pmatrix}
        H_z\pm E && H_x-i H_y
    \end{pmatrix}
\end{equation}
where $E=\sqrt{H_x^2+H_y^2+H_z^2}$. For an initially spin-down configuration, $c_\pm$ are proportional to
\begin{equation}
    \begin{split}
        c_\pm&\propto \begin{pmatrix}
        H_z\pm E && H_x-i H_y
    \end{pmatrix}\cdot \begin{pmatrix}
        \cos{\frac{Et}{\hbar}}-\frac{iH_z}{E}\sin{\frac{Et}{\hbar}} && -i\frac{H_x-iH_y}{E}\sin{\frac{Et}{\hbar}}\\
        -i\frac{H_x+iH_y}{E}\sin{\frac{Et}{\hbar}} &&  \cos{\frac{Et}{\hbar}}+\frac{iH_z}{E}\sin{\frac{Et}{\hbar}}
    \end{pmatrix}\cdot \begin{pmatrix}
        0 \\
       1
    \end{pmatrix}\\
    &=e^{\mp i\frac{\text{Re}[E]t}{\hbar}\pm\frac{\text{Im}[E]t}{\hbar}}(H_x-iH_y).
    \end{split}
\end{equation}
Therefore, under the normalization condition $|\psi|^2=|c_+|^2+|c_-|^2=1$, the overlap between the wavefunction with the right eigenstates are
\begin{equation}
    \begin{split}
        \frac{|c_\pm|^2}{|\psi|^2}&=\frac{|H_x-iH_y|^2 e^{\frac{\pm2\text{Im}[E]t}{\hbar}}}{|H_x-iH_y|^2\Big(e^{\frac{2\text{Im}[E]t}{\hbar}}+e^{-\frac{2\text{Im}[E]t}{\hbar}}\Big)}
        =\frac{e^{\pm\frac{2\text{Im}[E]t}{\hbar}}}{2\cosh{\frac{2\text{Im}[E]t}{\hbar}}}
    \end{split}
\end{equation}
hence we derive the equation used in the main text.

\end{widetext}
\bibliography{refs}

\end{document}